\documentclass{article} % For LaTeX2e
\usepackage{nips14submit_e,times}
\usepackage{hyperref}
\usepackage{url}
\usepackage{graphicx}
\usepackage{algorithm2e}
\usepackage{amsmath}
\DeclareMathOperator*{\argmin}{\arg\!\min}

\title{Use of Laplacian Projection Technique for Summarizing Likert Scale Annotations}

\author{
M. Iftekhar Tanveer\\
Electrical and Computer Engineering\\
University of Rochester\\
Rochester, NY 14627 \\
\texttt{itanveer@cs.rochester.edu}
}

\nipsfinalcopy % Uncomment for camera-ready version

\begin{document}

\maketitle

\begin{abstract}
  Summarizing Likert scale ratings from human annotators is an important step for collecting human judgments. In this project we study a novel, graph theoretic method for this purpose. We also analyze a few interesting properties for this approach using real annotation datasets.
\end{abstract}

\section{Introduction}
\label{sec:intro}
\begin{figure}[h]
\begin{center}
%\framebox[4.0in]{$\;$}
\includegraphics[width=0.5\linewidth]{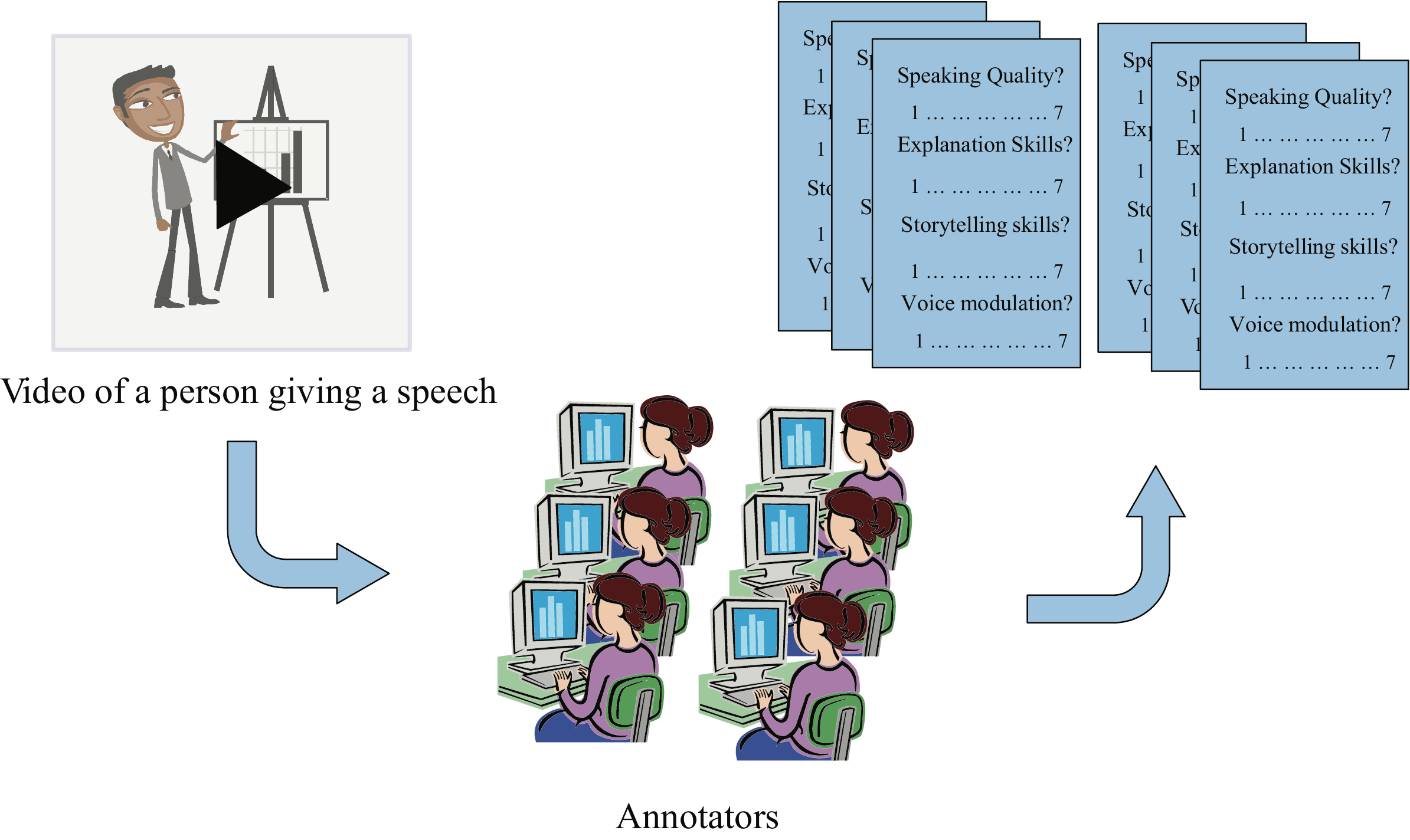}
\end{center}
\label{Fig:annotation}
\caption{Process of human annotation.}
\end{figure}

Likert scale is a popular method for quantifying and gathering human opinion. In disciplines like Behavioral Science, Psychology, or Human-Computer-Interaction, scientists use Likert scale to measure subjective opinions from human annotators. As human annotations are inherently noisy, it is often customary to collect the data from more than one annotators (Fig. \ref{Fig:annotation}). Traditionally, average of these ratings are computed to get a summary.

However, human annotators usually have their own bias on the ratings. For example, some annotators are biased towards high ratings, some towards low ratings; some annotators rate on a wide range, some others rate on a narrow range. As a result, computing averages without addressing these variations might lead to erroneous results. 

In classical literature on label denoising, researchers attempt to learn the underlying distribution of annotator ratings to rescale and compute the average. On the other hand, in this project we are interested to capture the neighborhood information from the datapoints. The rationale for this approach is as follows: Although humans have personal bias in assigning the exact values in the Likert scale, the idea of relative positions of the datapoints is universal. We employ a graph based technique to capture this neighborhood relationship within the datapoints.

\section{Literature Review}
In recent years, acquiring ground truth via accumulation of several unreliable crowd annotators is considered to be an important problem to address. An increasing body of literature is addressing this problem from various perspectives. For example, there are works involving complicated generative models \cite{Whitehill2009}\cite{Raykar2010} for denoising and aggregating crowd opinion. As these models involve latent variables, the solution approach typically involve Expectation Maximization (EM) \cite{Dempster1977} framework. These approaches are often criticized\cite{Karger2011} as EM is sensitive to initialization and can stick to local optima. Liu et al. \cite{Liu2012} approached this problem as an inference problem in probabilistic models. They used variational methods such as Belief Propagation and Mean Fields. 

However, none of these approaches consider the relative neighborhood of the vertices from graph perspective. As the annotators assign ratings to the subjects in comparison to one another; it should be reasonable to assume that the ground truth annotations captured in the relative distances among the datapoints.

\section{Problem Formulation}\label{sec:probfor}
\begin{figure}
\centering
\includegraphics[scale=0.5]{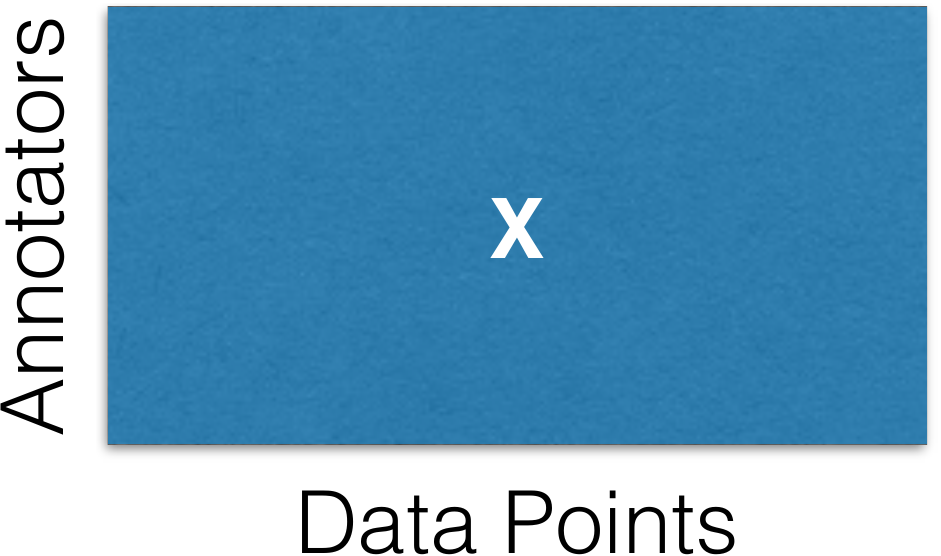}
\caption{The structure of the Annotation Matrix}
\label{Fig:dataStruct}
\end{figure}
Let us consider the annotation data is stored in an $m\times n$ matrix as shown in Fig.~\ref{Fig:dataStruct}. Each row of the matrix represents a unique annotator and each column represents a unique datapoint. The ratings are given in a K-point Likert scale. The goal of this problem is to formulate a linear embedding that projects the $m$ dimensional data points on to a one dimensional line while preserving the relative neighborhood. This idea is inspired from the concept of Locality Preserving Projection proposed by Niyogi et al.~\cite{Niyogi2004}. If $\mathbf{a}$ be the intended embedding, then the projection of datapoint $X_{:j}$ is $y_j$, which can be written as in Eq.~\eqref{eq:1dEmbedding}. 
\begin{equation}
y_j = \mathbf{a}^T X_{:j} \implies \mathbf{y}^T = \mathbf{a}^T X
\label{eq:1dEmbedding}
\end{equation}

In order to capture the neighborhood structure among the datapoints, we formulate a graph for each individual annotator. In the graph each node corresponds to a datapoint. In order to avoid encoding the subjective biases, we take the following simplest rule for forming the edges --- two nodes will be connected by an edge if and only if the two datapoints receive identical nonzero rating.  That is,
\begin{equation}
	w_{ij}^{(m)} = 
	\begin{cases}
		1 & \text{if node i and j has nonzero identical score by annotator m} \\
   		0 & \text{otherwise}.
	\end{cases}
	\label{eq:graphweight}
\end{equation}

This weight assignment captures the proximity of the datapoints in a higher dimensional space. The more annotators agree with this proximity information, the closer the datapoints are considered to be. To capture this structure, we formulate the final neighborhood graph by averaging the proximity weights $w_{ij}^{(m)}$ between nodes $i$ and $j$ for all the annotators. Now we project the datapoints on a 1D space so that the distances among the projections preserve the neighborhood structure composed by all the $w$'s. Mathematically we want to minimize the following. 
\begin{equation}	
	\begin{split}
		 & \frac{1}{M}\sum_{m=1}^{M}\sum_{(i,j)\in E_m} w_{ij}^{(m)}(y_i - y_j)^2 \\
		= & \frac{1}{M}\sum_{m=1}^{M}\mathbf{y}^T L_m \mathbf{y}\\
		= & \mathbf{y}^T \left( \frac{1}{M}\sum_{m=1}^{M} L_m \right) \mathbf{y}\\
		= & \mathbf{a}^T X L_{\text{avg}} X^T \mathbf{a}
		\label{eq:ObjFun}
	\end{split}
\end{equation}
where, $L_m := D_m - A_m$ is called the graph Laplacian for $m^{\text{th}}$ annotator graph. $D_m$ and $A_m$ are the degree matrix and adjacency matrix respectively for the $m^{\text{th}}$ annotator graph. As the graphs are undirected, so $A_m$ is symmetric. $D_m$ is obtained by performing row or column-wise sum and then placing the sums in the diagonal. As the degree of the nodes capture a natural measure of node importance, we impose a constraint $\mathbf{a}^TX D_{\text{avg}} X^T \mathbf{a} = 1$. Here $D_{\text{avg}}$ is the average degree matrix for all the annotators. Therefore, the optimization problem becomes:
\begin{equation}
\begin{split}
\argmin_{\mathbf{a}} \quad & \mathbf{a}^T X L_{\text{avg}} X^T \mathbf{a}\\
s.t. \quad & \mathbf{a}^TX D_{\text{avg}} X^T \mathbf{a} = 1
\end{split}
\label{eq:Optimization}
\end{equation}

\section{Optimization}\label{sec:opt}
We use Lagrange multiplier to construct the objective function as follows:
\begin{equation}
\mathcal{L}(\mathbf{a}) =  \mathbf{a}^T X L_{\text{avg}} X^T \mathbf{a} + \lambda (1 - \mathbf{a}^TX D_{\text{avg}} X^T \mathbf{a}).
\label{eq:opjFun2}
\end{equation} 
As both the Laplacian matrix and Degree matrix are positive semi-definite, $\mathcal{L}$ is a convex function of $\mathbf{a}$. Therefore, differentiating $\mathcal{L}$ with respect to $\mathbf{a}$ and setting to zero we get,  
\begin{equation}
\frac{\partial \mathcal{L}}{\partial \mathbf{a}} =  2X L_{\text{avg}} X^T \mathbf{a} - 2\lambda X D_{\text{avg}} X^T \mathbf{a}  = 0 \implies X L_{\text{avg}} X^T \mathbf{a} = \lambda X D_{\text{avg}} X^T \mathbf{a}.
\label{eq:opjFun3}
\end{equation}
Therefore $\mathbf{a}$ can be obtaining by solving the generalized Eigenvalue problem as shown in Eq.~\eqref{eq:opjFun3}. In order to get the solution that minimizes $\mathcal{L}$, we take the Eigenvector with smallest corresponding Eigenvalue.

\section{Algorithm}
From the discussions in Section~\ref{sec:probfor} and Section~\ref{sec:opt}, we can formulate an algorithm for computing the intended 1D embedding. We show it in Algorithm~\ref{algo:denoise}.
\begin{algorithm}[t]
 \caption{Algorithm for denoising the manual annotation using graph theory}\label{algo:denoise}
 \KwIn{Annotation Matrix, $X$}
 \KwOut{Denoised Annotations, $\mathbf{y}$}
  \textbf{Initialize:}\;
\ForEach {\textbf{annotator} m}{   
  Construct a Graph with Adjacency Matrix $A_m$ where each node represents a datapoint and edges satisfy Eq.~\eqref{eq:graphweight}\;
  }
  $A_{\text{avg}} \leftarrow \frac{1}{M}\sum_{m=1}^{M} A_m$\;
  Construct a diagonal matrix, $D_{\text{avg}}$, with entries $\sum_{\text{rows}}A_{\text{avg}}$\;
  $L:= D_{\text{avg}} - A_{\text{avg}}$\;
  Calculate Generalized Eigenvalue solution for Eq.~\eqref{eq:opjFun3}\;
  Return normalized Eigenvector with smallest corresponding Eigenvalue\;
\end{algorithm}

\section{Data}
\begin{figure}
\centering
\includegraphics[width=0.45\linewidth]{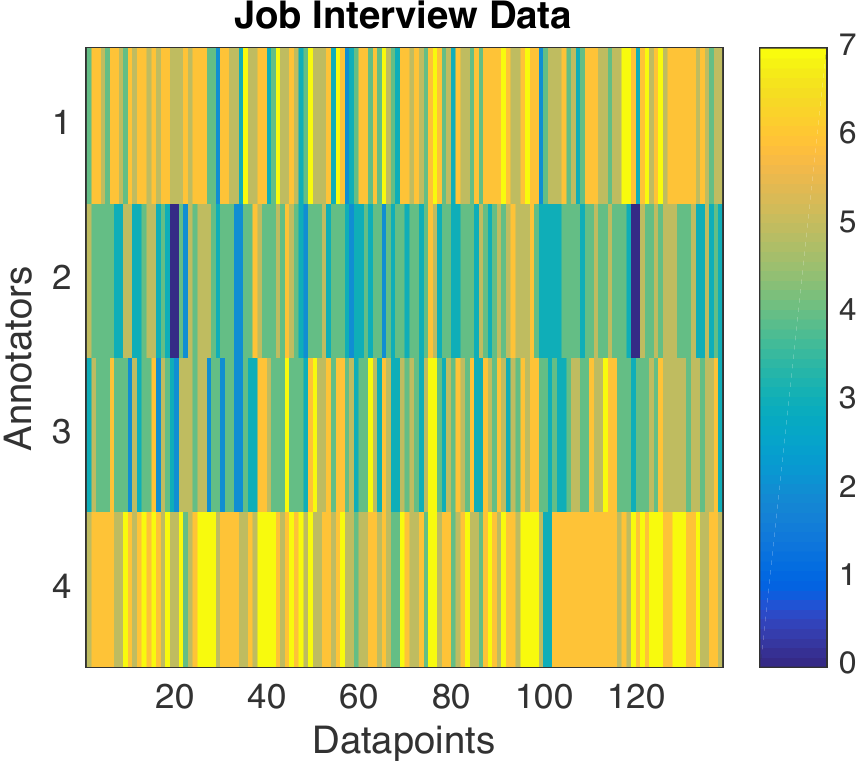}
\includegraphics[width=0.45\linewidth]{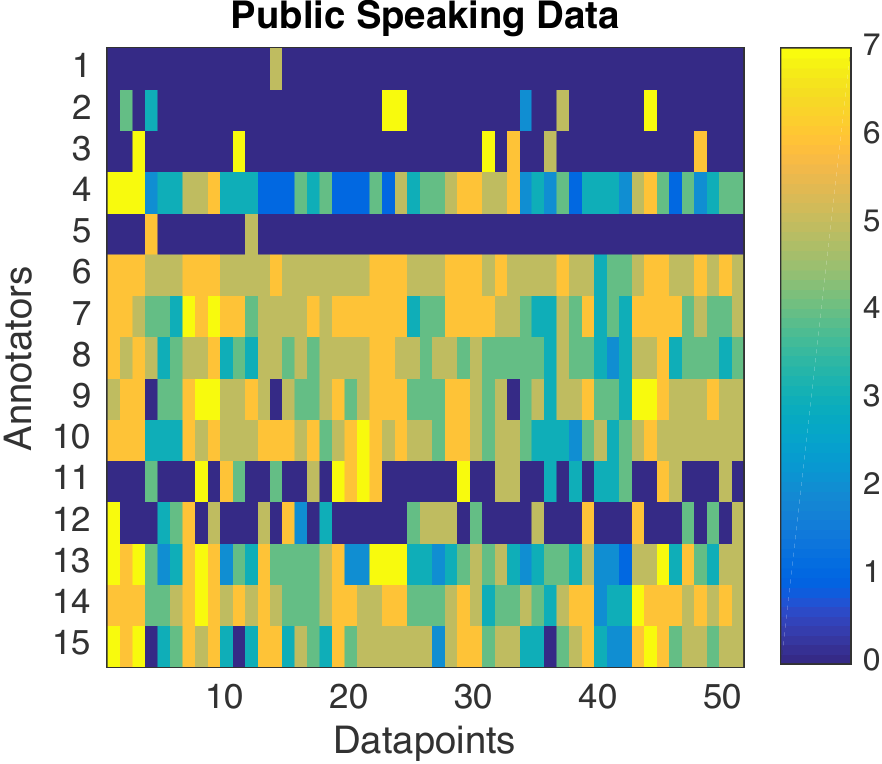}\caption{Job Interview Dataset (Left) and Public speaking Dataset (Right)}\label{Fig:datasets}
\end{figure}

We apply the algorithm on the following two datasets:
\begin{itemize}
\item Job interview dataset~\cite{Hoque2013}, and
\item Public speaking dataset~\cite{Tanveer2015}.
\end{itemize}
We used the annotators' response on the overall performance of the study participants. The data matrix, $X$, for both datasets are shown in Fig.~\ref{Fig:datasets}. There are 4 annotators and 138 datapoints (participants) in the job interview dataset. It is particularly evident from the picture of job interview dataset that the annotator 1 and 4 have a tendency to give higher rating than annotator 2 and 3.  In the public speaking dataset, there are 15 annotators and 51 datapoints. However, there are a number of missing values (rating is zero) in the public speaking dataset. 

\section{Results}
\begin{figure}
\centering
\includegraphics[width=\linewidth]{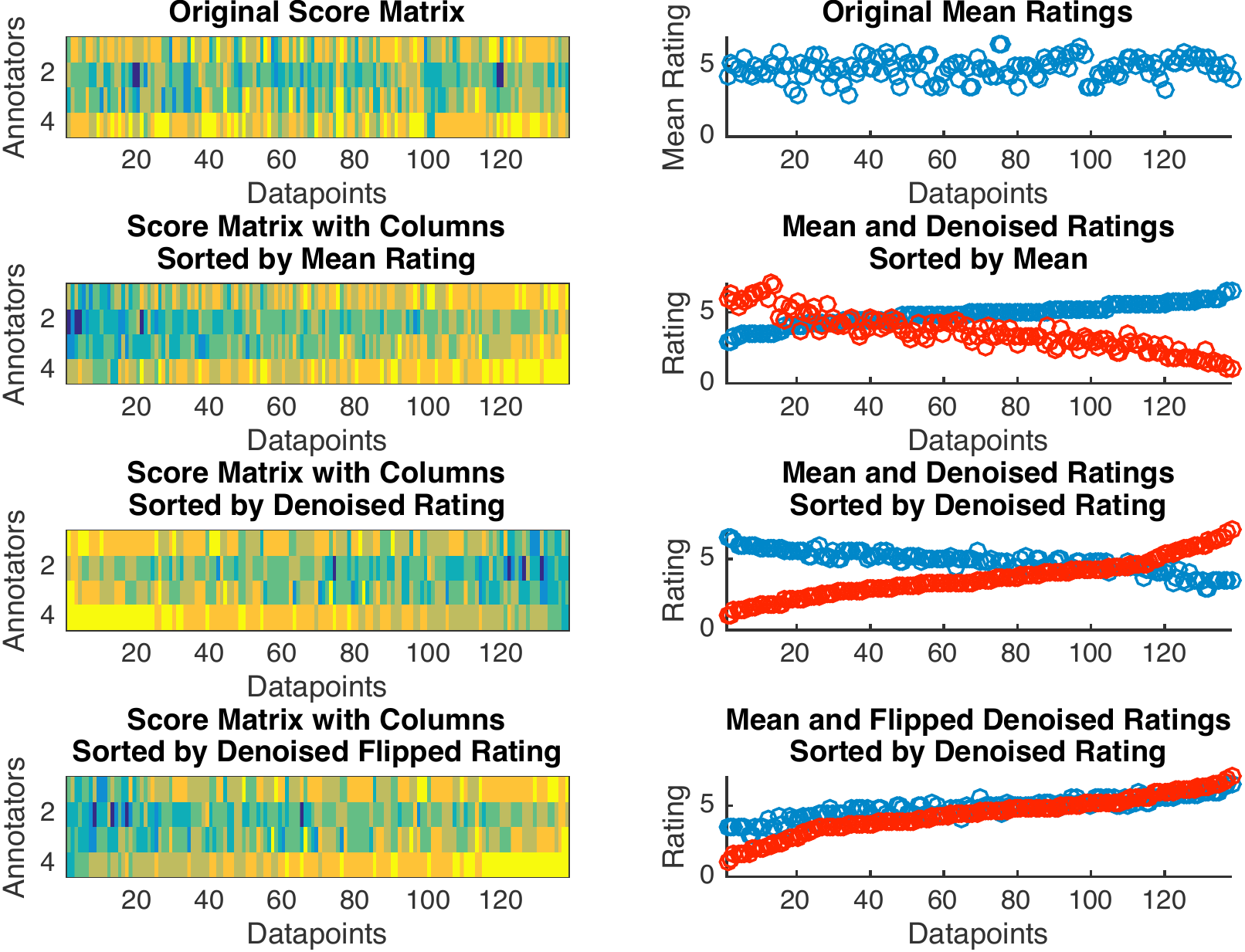}
\caption{Results of denoising for Job Interview Dataset}\label{Fig:jint_result}
\end{figure}

The results for the job interview dataset are shown in Fig.~\ref{Fig:jint_result}. The topmost row shows the original data matrix ($X$) on the left and the average of annotators' scores on the right.

In the second row, we sort the datapoints in ascending order of the mean ratings for representational convenience. The columns of the data matrix is also sorted accordingly. We also plot the denoised values of the ratings (i.e. the projected values obtained by the proposed algorithm) in red markers. It is interesting to notice that the projected values follow a flipped sequence than the mean ratings. This is due to the fact that our proposed optimization algorithm selects an embedding by only preserving the relative neighborhood among the datapoints. It does not preserve the absolute values. Consequently, the datapoints might arbitrarily take positive or negatively correlation with the mean values. However, in practice, it is not of a big concern as we can always flip the sequence by subtracting all the values in the sequence from the maximum allowable rating.

The third row represents the data matrix and the accumulated plots which are sorted based on the denoised ratings. If we compare the scatter plots (right hand side plot) of the second row with the third one, it is evident that the denoised ratings give finer discrimination among the datapoints than the mean values. As there are only four annotators, the means cannot discriminate among the datapoints less than $\frac{1}{4}^{\text{th}}$ of the ratings. As a result, significantly more datapoints receive same rating (notice the blue line on the right hand plot of the second row). However, as the denoised ratings consider the neighborhoods of the datapoints, it can discriminate with finer detail (notice the red line on the right hand plot of third row).

In the fourth row, the denoised ratings are flipped by subtracting them from the maximum rating, 7. This made the denoised ratings to be positively correlated with the mean values of the ratings. The datapoints were resorted in ascending order of the flipped denoised ratings. 

\begin{figure}
\centering
\includegraphics[width=\linewidth]{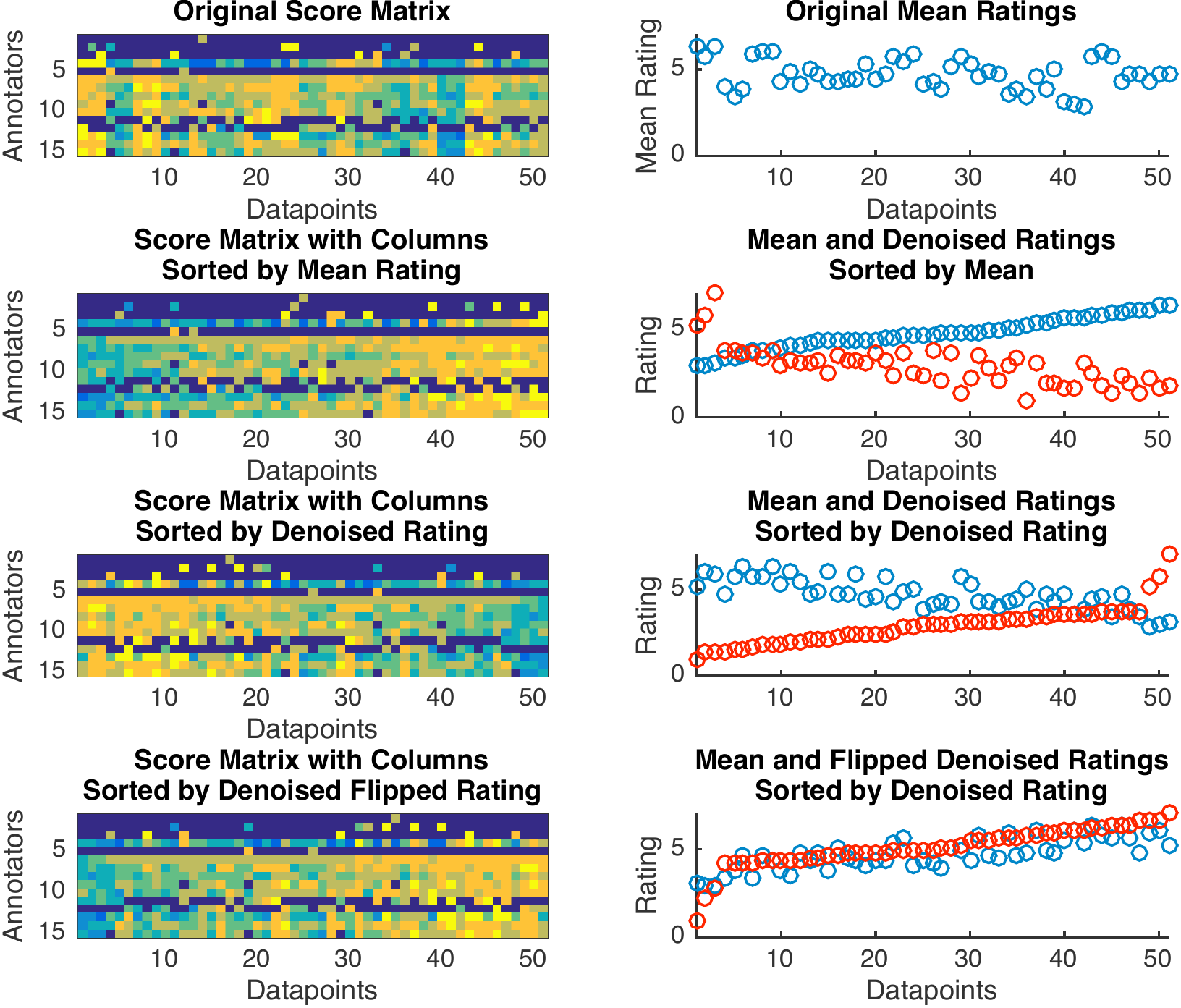}
\caption{Results of denoising for Public Speaking Dataset}\label{Fig:pspeak_result}
\end{figure}

In Fig.~\ref{Fig:pspeak_result}, we show similar results for public speaking dataset. While calculating the mean, we totally omitted the missing values so that they do not bias the mean ratings. In this dataset, the mean values are capable to show finer discrimination than the job interview dataset. This is due to the higher number of annotators. An interesting phenomenon is that the denoised ratings particularly enhanced the poor quality of three datapoints (notice the three leftmost points in the red line on 2nd row) while the mean ratings ``smooth out'' the differences. 

\section{Future Work and Conclusion}
In this project, we proposed a novel technique for summarizing the annotators opinion. The technique employs graph structure which captures the relative neighborhood among the datapoints. We applied the techniques on two datasets and compared with simple mean ratings.

In future, we will try to apply this technique with ground truth information of subjective data. Having the ground truth will enable us to better quantify the quality of this metric.

\bibliographystyle{abbrv}
\bibliography{project}

\end{document}